\documentclass[aps,prd,twocolumn,showkeys,amsmath,amssymb]{revtex4}
\usepackage{epsfig} 
\usepackage{amsmath}\allowdisplaybreaks
\usepackage{makecell}
\usepackage{array}
\usepackage{subfigure}
\usepackage{bm}
\usepackage{longtable}
\usepackage{tabularx}
\usepackage{booktabs}
\usepackage{appendix}
\usepackage{caption2}
\usepackage{amsfonts}
\usepackage{amsmath}
\usepackage{graphicx}
\graphicspath{{figs/}}
\usepackage{subfigure}
\usepackage{dcolumn}
\usepackage{siunitx}
\usepackage{bm}
\usepackage{booktabs}
\usepackage[utf8]{inputenc}
\usepackage{float}
\usepackage{longtable,lscape}
\usepackage{txfonts}
\usepackage{overpic}
\usepackage{amssymb}
\usepackage{indentfirst}
\usepackage{epsfig}
\usepackage{feynmf}   
\usepackage{epstopdf}   
\usepackage{slashed}  
\usepackage{multirow}
\usepackage{natbib}

\usepackage[colorlinks, citecolor=blue,anchorcolor=red,menucolor=red, linkcolor=red,filecolor=red,runcolor=red,urlcolor=blue,frenchlinks=red]{hyperref}
\usepackage{xcolor}
\usepackage{ulem}
\usepackage{color}

\makeatletter
\newcommand{\figcaption}{\def\@captype{figure}\caption}
\newcommand{\tabcaption}{\def\@captype{table}\caption}

\newcommand{\Rmnum}[1]{\expandafter\@slowromancap\romannumeral #1@}

\def\hlinewd#1{%
  \noalign{\ifnum0=`}\fi\hrule \@height #1 \futurelet
   \reserved@a\@xhline}
\makeatother

\newcommand\dqq{\Big\langle \bar q q \Big\rangle}

\newcommand\dGG{\Big\langle \alpha_{s} GG \Big\rangle}
\newcommand\dGGG{\Big\langle g_{s}^{3} G^{3} \Big\rangle}
\newcommand\dJJ{\Big\langle g_{s}^{4} jj \Big\rangle}

\begin{document}

\title{Investigations on heavy quarkonium hybrid mesons with exotic quantum numbers $J^{PC}=2^{+-}$}

\author{Qi-Nan Wang$^{1}$}
\email{wangqinan@bhu.edu.cn}

\author{Ding-Kun Lian$^2$}
\email{liandk@seu.edu.cn}
\author{Hua-Xing Chen$^2$}
\email{hxchen@seu.edu.cn}
\author{Wei Chen$^{3,\, 4}$}
\email{chenwei29@mail.sysu.edu.cn}
\affiliation{$^1$College of Physical Science and Technology, Bohai University, Jinzhou 121013, China\\
$^2$School of Physics, Southeast University, Nanjing 210094, China\\
$^3$School of Physics, Sun Yat-Sen University, Guangzhou 510275, China\\
$^4$Southern Center for Nuclear-Science Theory (SCNT), Institute of Modern Physics, 
Chinese Academy of Sciences, Huizhou 516000, Guangdong Province, China}

\begin{abstract}
We investigate the heavy quarkonium hybrid mesons with exotic quantum numbers $J^{PC}=2^{+-}$ via QCD sum rule method. We construct the currents with three Lorentz indices and calculate the correlation functions up to dimension six at the leading order of $\alpha _{s}$. The states with $J^{PC}=2^{+-}$ are extracted by constructing the corresponding projection operators. The obtained results indicate that the masses of $2^{+-}$ $\bar{c}Gc$ and $\bar{b}Gb$ hybrid states are about $4.49~\mathrm{GeV}$ and $10.48~\mathrm{GeV}$, respectively. We suggest to search for $\bar{c}Gc$ hybrid meson with $J^{PC}=2^{+-}$ in $J/\psi f_{0,1,2}$ and $\chi_{c0,1,2} \omega$ final states.
\end{abstract}

\keywords{hybrid meson, exotic quantum number, QCD sum rules, heavy flavor state }
\maketitle

\section{Introduction}

Quantum chromodynamics (QCD), as the established theory of the strong interaction, predicts exotic hadron states that lie outside the conventional quark model's description of $\bar{q}q$ mesons and $qqq$ baryons. Hybrid mesons ($\bar{q}gq$), which consist of a quark-antiquark pair accompanied by a valence gluon, are a prime example. The inclusion of explicit gluonic degrees of freedom allows these hybrids to exhibit quantum number combinations, such as $J^{PC}=0^{\pm-}$, $1^{-+}$, and $2^{+-}$, that are prohibited for traditional mesons.

Despite extensive research over many years, no hybrid meson has been definitively established experimentally. The leading candidates in the light-quark sector are the exotic isovector states $\pi_{1}(1400)$~\cite{IHEP-Brussels-LosAlamos-AnnecyLAPP:1988iqi}, $\pi_{1}(1600)$~\cite{E852:2001ikk}, and $\pi_{1}(2015)$~\cite{E852:2004gpn}, all with $I^GJ^{PC}=1^-1^{-+}$, along with the exotic isoscalar state $\eta_{1}(1855)$ with $I^GJ^{PC}=0^+1^{-+}$ ~\cite{BESIII:2022iwi,BESIII:2022riz}. Recent coupled-channel amplitude analyses~\cite{JPAC:2018zyd} of data from the $\eta^{(\prime)}\pi$ system~\cite{COMPASS:2014vkj} suggest that the $\pi_{1}(1400)$ and $\pi_{1}(1600)$ might correspond to a single state. Furthermore, it has been proposed that the $\eta_{1}(1855)$ could serve as the isoscalar partner of the $\pi_{1}(1600)$ within a hybrid nonet framework~\cite{Chen:2023ukh,Qiu:2022ktc,Shastry:2022upd,Shastry:2022mhk}. In the heavy-flavor sector, although numerous new $X$, $Y$, and $Z$ states have been discovered, no candidates with definitively exotic quantum numbers have been observed.

Theoretical predictions for the masses of hybrid mesons are crucial for understanding their properties. A variety of approaches have been employed, including Lattice QCD (LQCD)~\cite{Lacock:1996vy,Lacock:1997an,HadronSpectrum:2012gic,Ma:2020bex,Dudek:2011bn,Dudek:2009kk,Dudek:2013yja,Dudek:2009qf,Dudek:2010wm,Dudek:2011tt}, flux tube models~\cite{Isgur:1984bm,Isgur:1985vy,Burns:2006wz}, solutions of the Bethe-Salpeter equation~\cite{Burden:2002ps,Burden:1996nh}, and QCD sum rules~\cite{Balitsky:1982ps,Latorre:1984kc,Govaerts:1983ka,Govaerts:1984bk,Govaerts:1984hc,Govaerts:1986pp,Balitsky:1986hf,Huang:2014hya,Barsbay:2022gtu,Chen:2021smz,Chen:2022qpd,Chen:2013eha,Ho:2018cat,Ho:2019org,Palameta:2018yce,Wang:2025ypo,Wang:2025sic}. For example, the MIT bag model~\cite{Barnes:1982tx,Chanowitz:1982qj} identifies the lightest hybrid supermultiplet containing states with $J^{PC}=1^{--}$, $0^{-+}$, $1^{-+}$, and $2^{-+}$, among which only the $1^{-+}$ is exotic. In this model, the $0^{+-}$ and $2^{+-}$ hybrids are assigned to higher-mass supermultiplets, while the $0^{--}$ state is predicted to be considerably heavier due to a different gluonic excitation pattern~\cite{HadronSpectrum:2012gic,Chen:2013zia,Meyer:2015eta}. This overall mass hierarchy is generally supported by LQCD calculations~\cite{Dudek:2011bn,Dudek:2009qf,Dudek:2010wm,Dudek:2011tt} and studies in Coulomb gauge QCD~\cite{Guo:2008yz}.

Historically, hybrid mesons with some exotic quantum numbers, for instance $J^{PC}=0^{+-}$ and $2^{+-}$, have received significantly less attention compared to the $1^{-+}$ case. This is primarily because no experimental signals have been observed. The PSS model~\cite{Page:1998gz} and a recent study \cite{Farina:2025zhm}, however, predict that $2^{+-}$ hybrid states will have exceptionally narrow widths. If this prediction is accurate, such states would become prominent and readily detectable in specific decay channels. This provides a compelling reason to conduct a dedicated investigation into hybrid mesons characterized by $J^{PC}=2^{+-}$ quantum numbers.

Some Lattice QCD calculations suggest masses of $J^{PC}=2^{+-}$ hybrids around $2.4-2.8 ~\mathrm{GeV}$ for light flavored hybrids~\cite{Dudek:2011bn,Dudek:2013yja}, $4.4-4.5~\mathrm{GeV}$ for $\bar{c}Gc$ hybrids~\cite{HadronSpectrum:2012gic} and $11.1~\mathrm{GeV}$ for $\bar{b}Gb$ hybrids~\cite{Ryan:2020iog}. Born-Oppenheimer effective field theory also give mass spectra of heavy quarkonium hybrids, including states with $J^{PC}=2^{+-}$~\cite{Soto:2023lbh}. In Ref.~\cite{Wang:2023whb}, we investigate the mass spectra of light hybrid mesons with $J^{PC}=2^{+-}$ in QCD sum rule method, and the obtained results are consistent with the LQCD calculations.

To further investigate the mass spectra of hybrid mesons, in this study, we employ QCD sum rules to investigate heavy quarkonium hybrid mesons with $J^{PC}=2^{+-}$ quantum numbers following Ref.~\cite{Wang:2023whb}. The structure of this letter is as follows: In Sec.~\ref{Sec:2}, we construct local hybrid interpolating currents without covariant derivative operators and construct the corresponding projection operators to extract the states with $J^{PC}=2^{+-}$. The sum rules of hadronic mass can be obtained in Sec.~\ref{Sec:3} and the masses of heavy quarkonium hybrid mesons are obtained in Sec.~\ref{Sec:4}. The last section is a brief discussion.

\section{Interpolating Currents and Projection Operators}\label{Sec:2}
A hybrid current is typically composed of a quark-antiquark pair and an excited gluon field. As described in Ref.~\cite{Govaerts:1986pp}, hybrid currents with two Lorentz indices may couple to states having $J^{PC}=2^{++}$ and $2^{-+}$. For states with exotic quantum numbers $J^{PC}=2^{+-}$, we construct interpolating hybrid currents that involve three Lorentz indices. Utilizing the gluon field strength tensor $G_{\mu \nu }(x)$ and Dirac matrices, the potential hybrid currents with three Lorentz indices can be formulated as
\begin{align}
  \label{Eq:current1} J_{\alpha \mu  \nu }^{1}&=\bar{Q} g_s \gamma _\alpha \gamma_5 G_{\mu \nu } Q, ~~~~J^{PC}=(0, \,1, \,2)^{\pm-}, \,\\
  \label{Eq:current2}  J_{\alpha \mu \nu  }^{2}&=\bar{Q} g_s \gamma_\alpha \gamma_5 \tilde{G}_{\mu \nu } Q, ~~~~J^{PC}=(0, \,1, \,2)^{\pm-}, \,
\end{align}
where $Q$ is a heavy quark field ($c$ or $b$), $g_s$ is the strong coupling constant which is defined via $D_{\mu}=\partial_{\mu}+i g_{s}A_{\mu}$, and $\tilde{G}_{\mu \nu }=\frac{1}{2}\varepsilon_{\mu \nu \alpha \beta }G^{\alpha \beta }$ is the dual gluon field strength. 

To investigate the physical states with $J^{PC}=2^{+-}$, we consider the  couplings between the interpolating current and different hadron states as follows
\begin{align}
\label{Eq:coupling1}
\left\langle 0\left|J_{\alpha \mu \nu}\right| 0^{(-P) C}(p)\right\rangle&=Z_1^0 p_\alpha g_{\mu \nu}+Z_2^0 p_\mu g_{\alpha \nu}+Z_3^0 p_\nu g_{\alpha \mu}\nonumber\\
&+Z_4^0 p_\alpha p_\mu p_\nu  , \,\\ \label{Eq:coupling2}
 \left\langle 0\left|J_{\alpha \mu \nu}\right| 0^{PC}(p)\right\rangle&=Z_5^0 \varepsilon_{\alpha \mu \nu \tau} p^\tau  , \,\\ \label{Eq:coupling3}
 \left\langle 0\left|J_{\alpha \mu \nu}\right| 1^{PC}(p)\right\rangle&=Z_1^1 \epsilon_\alpha g_{\mu \nu}+Z_2^1 \epsilon_\mu g_{\alpha \nu}+Z_3^1 \epsilon_\nu g_{\alpha \mu}\nonumber\\
 &+Z_4^1 \epsilon_\alpha p_\mu p_\nu+Z_5^1 \epsilon_\mu p_\alpha p_\nu+Z_6^1 \epsilon_\nu p_\alpha p_\mu  , \,\\ \label{Eq:coupling4}
 \left\langle 0\left|J_{\alpha \mu \nu}\right| 1^{(-P) C}(p)\right\rangle&=Z_7^1 \varepsilon_{\alpha \mu \nu \tau} \epsilon^\tau+Z_8^1 \varepsilon_{\alpha \mu \tau \lambda} \epsilon^\tau p^\lambda p_\nu\nonumber\\
 &+Z_9^1 \varepsilon_{\alpha \nu \tau \lambda} \epsilon^\tau p^\lambda p_\mu  , \,\\ \label{Eq:coupling5}
 \left\langle 0\left|J_{\alpha \mu \nu}\right| 2^{(-P)C}(p)\right\rangle&=Z_1^2 \epsilon_{\alpha \mu} p_\nu+Z_2^2 \epsilon_{\alpha \nu} p_\mu+Z_3^2 \epsilon_{\mu \nu} p_\alpha  , \,\\ \label{Eq:coupling6}
\left\langle 0\left|J_{\alpha \mu \nu}\right| 2^{PC}(p)\right\rangle&=Z_4^2 \varepsilon_{\alpha \mu \tau \theta} \epsilon_\nu^{~\tau} p^\theta+Z_5^2 \varepsilon_{\alpha \nu \tau \theta} \epsilon_\mu^{~\tau} p^\theta  , \,\\
 \left\langle 0\left|J_{\alpha \mu \nu}\right| 3^{PC}(p)\right\rangle&=Z_1^3 \epsilon_{\alpha \mu \nu} , \,
\label{Eq:coupling7}
\end{align}
where $\epsilon_{\alpha}$, $\epsilon_{\alpha\mu}$, $\epsilon_{\alpha\mu\nu }$ are the polarization tensors for the spin-1, spin-2, spin-3 states. $\varepsilon_{\alpha \mu \nu \tau}$ is the Levi-Civita tensor. The couplings in Eqs. (\ref{Eq:coupling1})-(\ref{Eq:coupling7}) represent the most general form of couplings for currents with three Lorentz indices and are independent of whether these indices exhibit symmetry or antisymmetry.
It should be noted that the interpolating currents in Eqs.~(\ref{Eq:current1}) and ~(\ref{Eq:current2}) can not couple to any spin-3 state since their last two Lorentz indices are antisymmetric while the spin-3 polarization tensor $\epsilon_{\alpha\mu\nu}$ is completely symmetric.

Since the parities for the currents $J_{\alpha \mu \nu }^{1}$, and $J_{\alpha \mu \nu }^{2}$ are opposite, they couple to the $2^{+-}$ hybrid states via different coupling relations in Eq.~\eqref{Eq:coupling6} and Eq.~\eqref{Eq:coupling5}, respectively. 
However, due to the antisymmetric properties under the exchange of $\mu$ and $\nu$ in the currents $J_{\alpha \mu \nu }^{1,2}$, we can rewrite the coupling in Eqs.~\eqref{Eq:coupling6} and \eqref{Eq:coupling5} in another way
\begin{align}
  \left\langle 0\left|J_{\alpha \mu \nu }^{1}\right| 2^{+-}(p)\right\rangle
   =&f_{2}\left(\varepsilon_{\alpha \mu  \tau \theta} \epsilon_\nu^{~\tau} p^\theta-\varepsilon_{\alpha \nu  \tau \theta} \epsilon_\mu ^{~\tau} p^\theta\right), \,
  \label{Eq:coupling71}
\end{align}
  \begin{align}
  \left\langle 0\left|J_{\alpha \mu \nu }^{2}\right| 2^{+-}(p)\right\rangle =&
  f_{2}^{\prime}(\epsilon_{\alpha \mu } p_\nu  -\epsilon_{\alpha \nu } p_\mu )\, .
  \label{Eq:coupling81}
\end{align}
In above couplings, the symmetric part under the exchange of $\mu$ and $\nu$ disappears  due to the antisymmetric properties for currents $J_{\alpha \mu \nu }^{1,2}$.
 
One can construct the normalized projection operator for the  $2^{+-}$ states from current $J_{\alpha \mu \nu }^{1}$
\begin{align}
  \mathbb{P}_{\alpha_{1}\mu_{1}\nu_{1},\alpha_{2}\mu_{2}\nu_{2}}=&\frac{1}{20p^{2}}
  \sum \left(\varepsilon_{\alpha_{1} \mu_{1}  \tau_{1}  \theta_{1} } \epsilon_{\nu_{1} }^{~\tau_{1} } p^{\theta _{1}}-\varepsilon_{\alpha_{1}  \nu_{1}  \tau_{1}  \theta_{1} } \epsilon_{\mu_{1} }^{~\tau_{1} } p^{\theta_{1} }\right)\nonumber\\&\times \left(\varepsilon_{\alpha_{2}  \mu_{2} \tau_{2} \theta_{2}} \epsilon_{\nu_{2}}^{~\tau_{2}*} p^{\theta_{2}}-\varepsilon_{\alpha_{2} \nu_{2} \tau_{2} \theta_{2}} \epsilon_{\mu_{2}}^{~\tau_{2}*} p^{\theta_{2}}\right)  , \,
  \label{Eq:Projector1}
\end{align}
where the summation over polarization of the tensor $\epsilon_{\alpha \beta}$ is
\begin{equation}
\sum \epsilon_{\alpha_{1} \beta_{1}}  \epsilon_{\alpha_{2} \beta_{2}}^{*} =\frac{1}{2}(\eta_{\alpha_{1} \alpha_{2}} \eta_{\beta_{1} \beta_{2}}+\eta_{\alpha_{1} \beta_{2}} \eta_{\beta_{1} \alpha_{2}}-\frac{2}{3} \eta_{\alpha_{1} \beta_{1}} \eta_{\alpha_{2} \beta_{2}}) , \,
\end{equation}
with
\begin{equation}
\eta_{\alpha\beta}=\frac{p_{\alpha} p_{\beta}}{p^{2}}-g_{\alpha \beta} \,,
\end{equation}
and from $J_{\alpha \mu \nu }^{2}$
\begin{align}
  \mathbb{P}_{\alpha_{1}\mu_{1}\nu_{1},\alpha_{2}\mu_{2}\nu_{2}}^{\prime}=&\frac{1}{20p^{2}}
  \sum \left(\epsilon_{\alpha_{1} \mu_{1} } p_{\nu_{1}}  -\epsilon_{\alpha_{1}  \nu_{1}  } p_{\mu_{1}} \right)\nonumber\\&\times\left(\epsilon_{\alpha_{2}  \mu_{2} }^{\ast} p_{\nu_{2}}  -\epsilon_{\alpha_{2} \nu_{2} }^{\ast} p_{\mu_{2}}\right)  \, .
  \label{Eq:Projector2}
\end{align}
It turns out that the $2^{+-}$ hybrid states obtained from the currents $J_{\alpha \mu \nu}^{1}$ are just the same states as those derived from $J_{\alpha \mu \nu}^{2}$. 
Therefore, in the subsequent investigations, we focus solely on the interpolating currents $J_{\alpha \mu \nu}^{1}$ to explore the mass spectra of the $2^{+-}$ hybrid states.

\section{Formalism of QCD sum rules}\label{Sec:3}
The two-point correlation functions of the current $J_{\alpha \mu \nu}^{1}$ in Eq.~(\ref{Eq:current1}) can be written as
\begin{equation}
\begin{aligned}
 \Pi_{\alpha_{1}\mu_{1} \nu_{1},\alpha_{2}\mu_{2} \nu_{2}}(p^{2}) &=i \int d^{4} x e^{i p \cdot x}\left\langle 0\left|T\left[J_{\alpha_{1}\mu_{1} \nu_{1}}^{1}(x) J_{\alpha_{2}\mu_{2} \nu_{2}}^{1\dagger}(0)\right]\right| 0\right\rangle \, .
 \label{Eq:correlator}
\end{aligned}
\end{equation}
We shall investigate the $J^{PC}=2^{+-}$ hybrid states in this work, which can be extracted by applying the projection operators defined in Eq.~\eqref{Eq:Projector1}
\begin{align}
    \Pi(p^{2}) &=\mathbb{P}^{\alpha_{1}\mu_{1}\nu_{1},\alpha_{2}\mu_{2}\nu_{2}} \Pi_{\alpha_{1}\mu_{1} \nu_{1},\alpha_{2}\mu_{2} \nu_{2}}(p^{2}) \, .
   \label{Eq:Pi_2}
\end{align}

At the hadronic level, the correlation function $\Pi(p^{2})$ can be usually described via the dispersion relation
\begin{equation}
\Pi(p^{2})=\frac{(p^{2})^{N}}{\pi} \int_{4m_{Q}^{2}}^{\infty} \frac{\operatorname{Im} \Pi(s)}{s^{N}\left(s-p^{2}-i \epsilon\right)} d s+\sum_{n=0}^{N-1} b_{n}(p^{2})^{n}, \,
\label{Cor-Spe}
\end{equation}
where the $b_n$ is the subtraction constant, and $N$ is the number of subtractions required to make the integral convergent. In QCD sum rules, the imaginary part of the correlation function is defined as the spectral function
\begin{equation}
\begin{aligned}
\rho (s)\equiv\frac{1}{\pi} \text{Im}\Pi(s)=f^{2}m_{H}^{2}\delta(s-m_{H}^{2})+\cdots, \,
\end{aligned}
\end{equation}
in which the “one pole plus continuum” parametrization assumption is used and “$\cdots$” represents the QCD continuum and higher states. The parameters $f$ and $m_{H}$ are the coupling constant and mass of the lowest-lying hadron state $H$, respectively. One should note that at higher energies where there can be more resonances appearing, the number density of states is an increasing function of the center of mass energy, which will reduce the rationality of this assumption.

To enhance the convergence of the OPE series and to minimize the impact of the continuum and excited states, the Borel transformation technique is applied to the correlation functions at both the hadronic and quark-gluon levels. Subsequently, the QCD sum rules are derived as
\begin{equation}
\Pi\left(M_{B}^{2},s_{0}\right)=f^{2} m_{H}^{2}e^{-m_{H}^{2} / M_{B}^{2}}=\int_{4m_{Q}^{2}}^{s_{0}} d s e^{-s / M_{B}^{2}} \rho(s), \,
\end{equation}
where $M_{B}^{2}$ is the squared Borel mass introduced via the Borel transformation and $s_0$ is the continuum threshold.
Then the hadron mass of the lowest-lying hybrid state can be extracted as
\begin{align}
   m_{H}\left(M_{B}^{2},s_{0}\right) & =\sqrt{\frac{\frac{\partial}{\partial\left(-1 / M_B^2\right)} \Pi\left(M_{B}^{2},s_{0}\right)}{\Pi\left(M_{B}^{2},s_{0}\right)}} \,.
\end{align}

Since the Borel mass is an intermediate parameter, it should not be relevant to the physical state, i.e., to obtain the optimal value of the continuum threshold $s_0$, the variation of the extracted hadron mass $m$ with respect to $M_{B}^{2}$ should be minimized. The parameters $M_{B}^{2}$ and $s_{0}$ in the QCD sum rule analyses can be determined by requiring a suitable OPE convergence which in this work can be defined as
\begin{equation}
  R_{D=6}=\left|\frac{\Pi^{D=6}\left(M_{B}^{2}, \infty\right)}{\Pi^{t o t}\left(M_{B}^{2}, \infty\right)}\right| , \,
  \end{equation}
and a big enough pole contribution(P.C) which can also be defined as
\begin{equation}
  \text{P.C}=\frac{\Pi\left(M_{B}^{2}, s_0\right)}{\Pi\left(M_{B}^{2}, \infty\right)} \, .
\end{equation}
We calculate the correlation functions for hybrid states with quantum numbers $J^{PC}=2^{+-}$, including condensates up to dimension 6. Contributions from operators of higher dimensions are negligible and are thus omitted. The detailed expressions, which are somewhat intricate, are provided in the Appendix.~\ref{appendix1}. 

\section{Numerical analyses and mass predictions}\label{Sec:4}
In this section, we perform the QCD sum rule analyses for the heavy quarkonium hybrid states with $J^{PC}=2^{+-}$. We use the following values for various QCD parameters~\cite{Jamin:2002ev,Narison:2011xe,Narison:2018dcr,ParticleDataGroup:2024cfk}.
\begin{align}
  \alpha_{s}(\mu)&=\frac{4\pi}{\left(11-\frac{2n_{f}}{3}\right)\ln{(\mu^{2}/\Lambda_{QCD}^{2}})}, \,\nonumber\\
  m_{c}(\overline{m} _{c}) & = 1.27  ~\mathrm{GeV}, \,\nonumber\\
  m_{b}(\overline{m} _{b}) & = 4.18\pm 0.01  ~\mathrm{GeV}, \,\nonumber\\
  \dqq & =-(0.24 \pm 0.01)^3 ~\mathrm{GeV}^3 , \, \\
  \dGG & =(6.35  \pm 0.35) \times 10^{-2} ~\mathrm{GeV}^4 , \,\nonumber\\
  \dGGG & =-(8.2  \pm 1.0) \times \dGG ~\mathrm{GeV}^2, \,\nonumber\\
  \left\langle g_s^4 j j\right\rangle&=-\frac{4}{3} g_s^4\langle\bar{q} q\rangle^2 \, .\nonumber
\end{align}
The condensate values listed above are taken at energy renormalization scale $\mu=1\,\mathrm{GeV}$. To maintain energy renormalization scale consistency, we use renormalization group to choose the condensates and masses of heavy quarks at $\mu=\overline{m} _{c}$ and $\overline{m} _{b}$ for $\bar{c}Gc$ and $\bar{b}Gb$ system, respectively. $\Lambda_{QCD}$ is set to be $210 ~\mathrm{MeV}$ and $292 ~\mathrm{MeV}$ for $n_{f}=5$ and $4$, respectively. 

\begin{figure}[h!!]
  \centering
  \includegraphics[width=5cm]{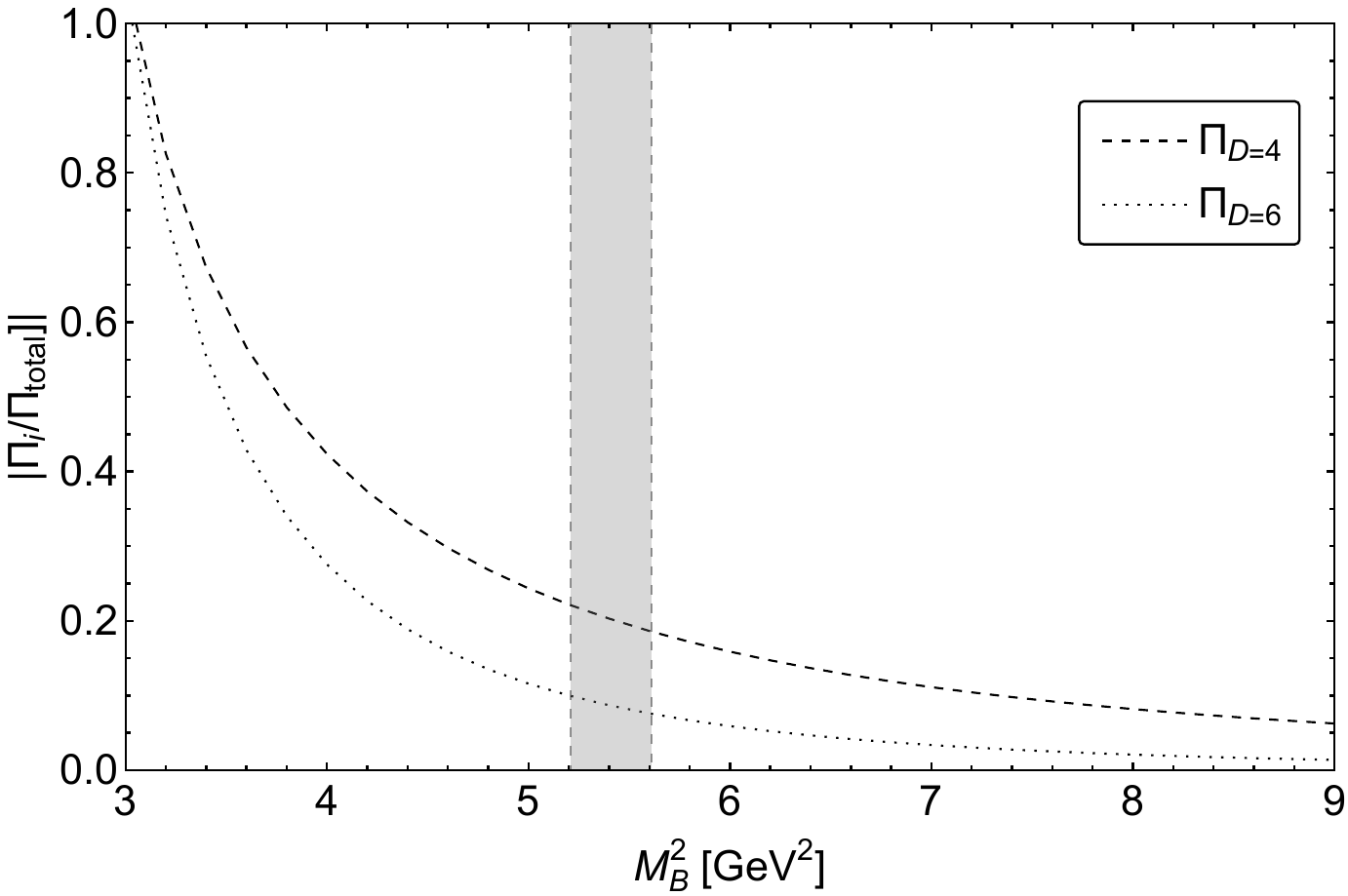}\\
  \caption{OPE convergence for the $\bar{c}Gc$ hybrid state with $J^{PC}=2^{+-}$.}
\label{fig:Convergence-cGc}
\end{figure}

We firstly take the $\bar{c}Gc$ hybrid state as an example to show the numerical analysis. To maintain the stability of the OPE series, it is essential that the Borel parameter $M_{B}^{2}$ is sufficiently large. Our requirement is that the contributions from dimension-6 condensates must not surpass 10\% of the total contribution, thereby setting a lower bound for $M_{B}^{2}$ at $5.21~\mathrm{GeV}^{2}$. The proportion of contributions from various condensates is illustrated in Fig.~\ref{fig:Convergence-cGc}, which confirms the satisfactory convergence behavior of the OPE series.

\begin{figure}[h!!]
  \centering
  \subfigure[]{\includegraphics[width=4.2cm]{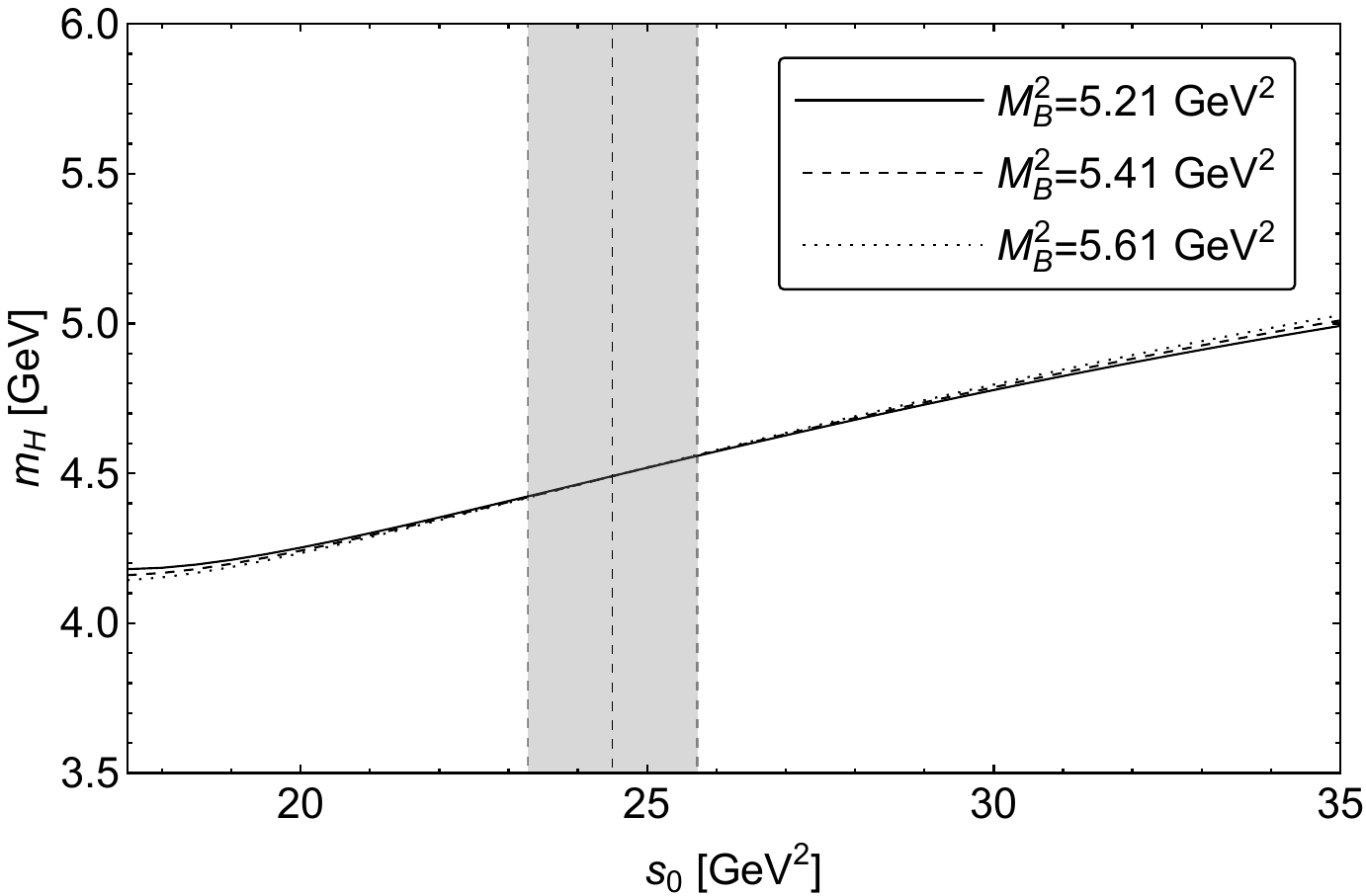}\label{fig:s0-mH-cGc}}
  \subfigure[]{\includegraphics[width=4.2cm]{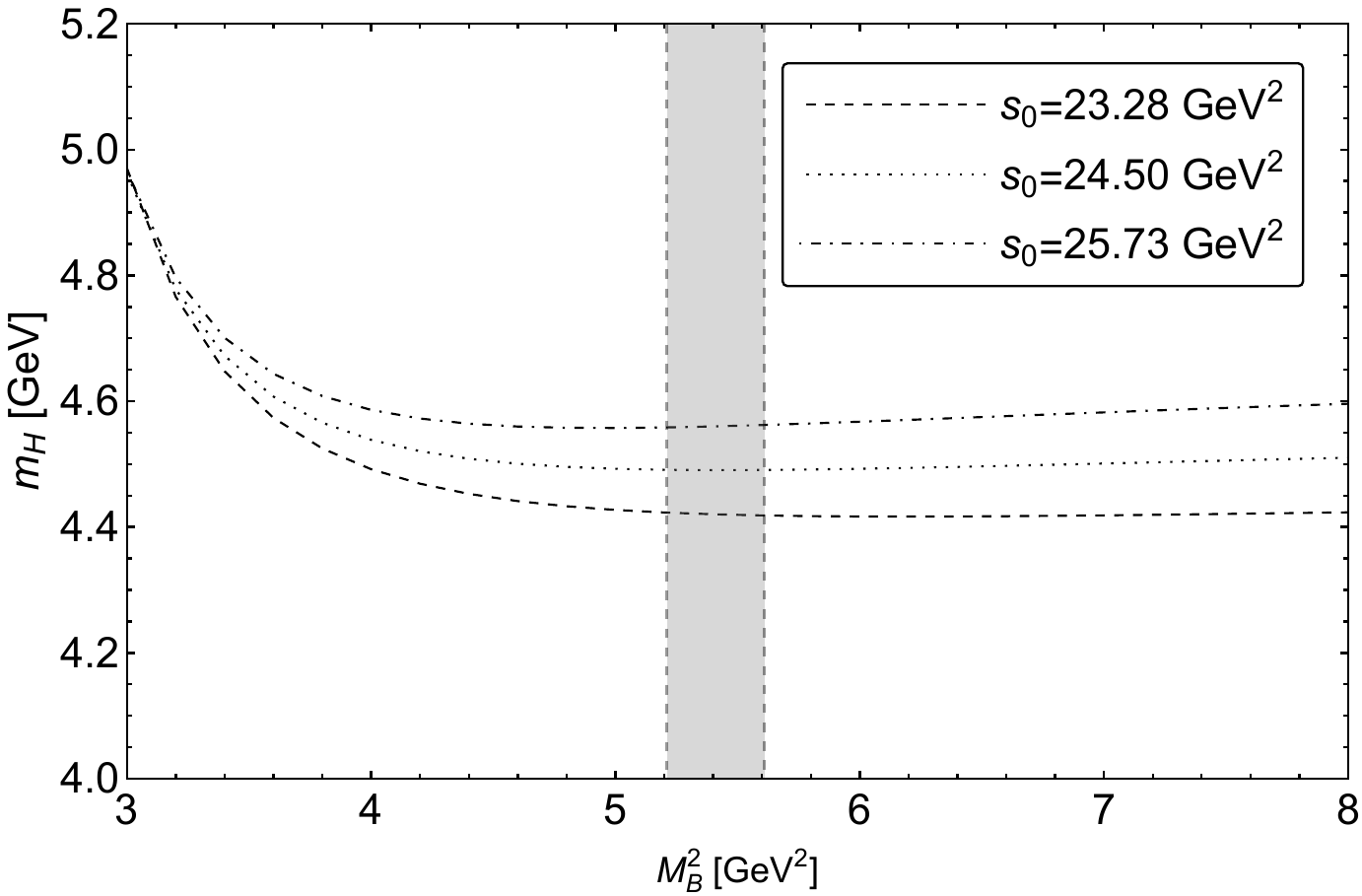}\label{fig:MB-mH-cGc}}\\
  \caption{Variation of $m_{H}$ with (a)$s_{0}$ and (b)$M_{B}^{2}$ corresponding to the $J^{PC}=2^{+-}$ $\bar{c}Gc$ hybrid state.}
\label{fig:Result-cGc}
\end{figure}

To ascertain the upper limit of $M_{B}^{2}$, we first determine the value of $s_{0}$. As discussed above, the hadron mass $m_{H}$ should ideally be independent of the Borel parameter $M_{B}^{2}$. Fig.~\ref{fig:s0-mH-cGc} shows the relationship between $m_{H}$ and the continuum threshold $s_{0}$ for different values of $M_{B}^{2}$. The analysis reveals that the dependence of $m_{H}$ on $M_{B}^{2}$ is minimized around $s_{0}=24.50~\mathrm{GeV}^{2}$. Consequently, the feasible range for $s_{0}$ is established as $23.28 \leq s_0 \leq 25.73~\mathrm{GeV}^{2}$, with a 5\% uncertainty assigned to $s_{0}$ here. The upper limit of $M_{B}^{2}$ is then determined by requiring that the P.C exceeds 30\%. Ultimately, the working range for the Borel parameter is set to $5.21 \leq M_{B}^{2} \leq 5.61~\mathrm{GeV}^{2}$. We present the Borel plots within the specified parameter regions in Fig.~\ref{fig:MB-mH-cGc}, where the stability of the QCD sum rules is adequate to reliably predict the hadron mass of the $2^{+-}$ $\bar{c}Gc$ hybrid state
\begin{align}
    m_{\bar{c}Gc}&=4.49_{-0.08}^{+0.08}~\mathrm{GeV}, \,
\end{align}
where the errors are mainly from the uncertainties of the continuum threshold $s_{0}$, and various QCD condensates. The error from the Borel mass is small enough to be neglected. The result obtained above is consistent with the Lattice QCD calculations in Ref.~\cite{HadronSpectrum:2012gic}.

\begin{figure}[h!!]
  \centering
  \includegraphics[width=5cm]{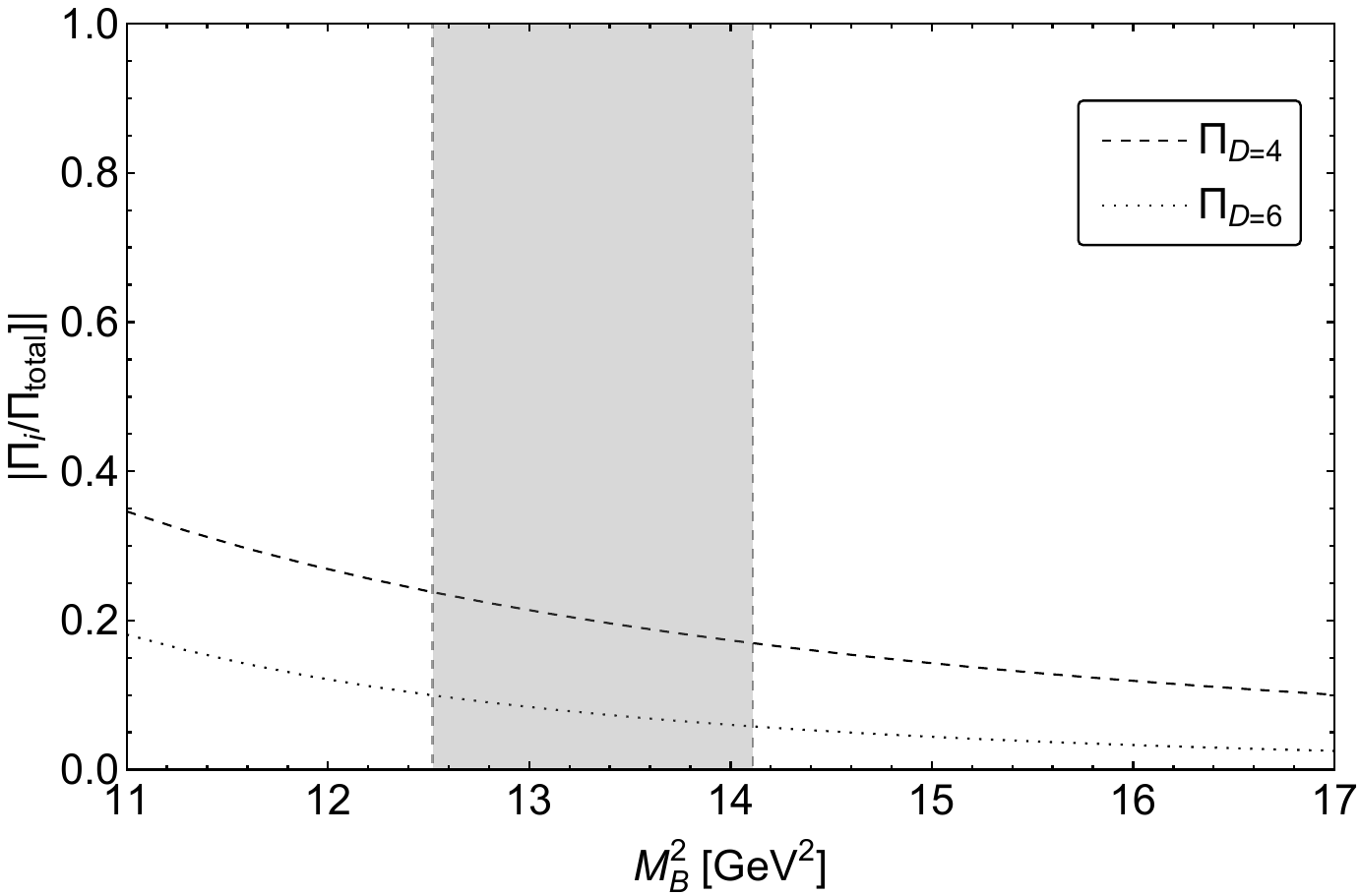}\\
  \caption{OPE convergence for the $\bar{b}Gb$ hybrid  state with $J^{PC}=2^{+-}$.}
\label{fig:Convergence-bGb}
\end{figure}

\begin{figure}[h!!]
  \centering
  \subfigure[]{\includegraphics[width=4.2cm]{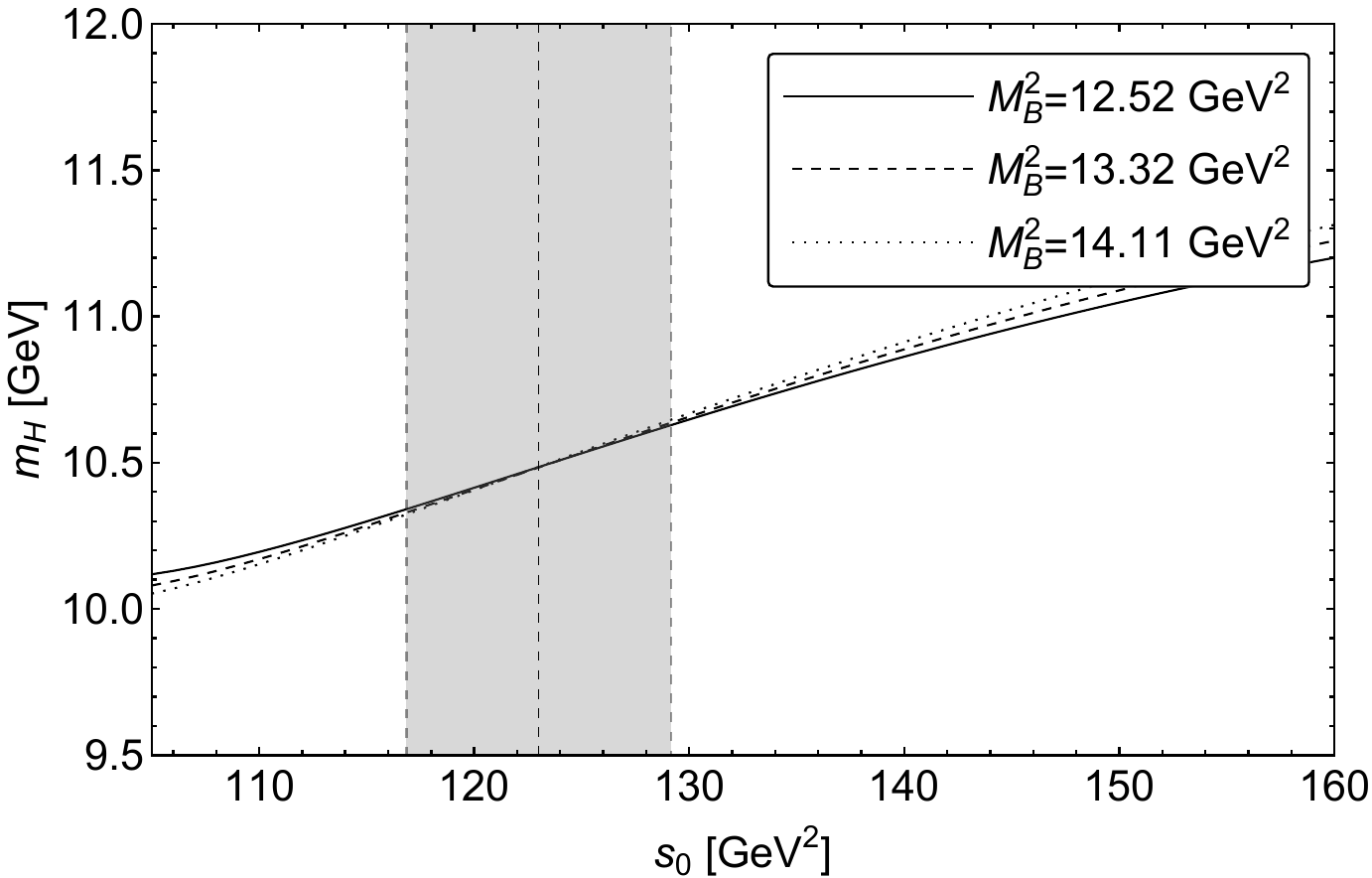}\label{fig:s0-mH-bGb}}
  \subfigure[]{\includegraphics[width=4.2cm]{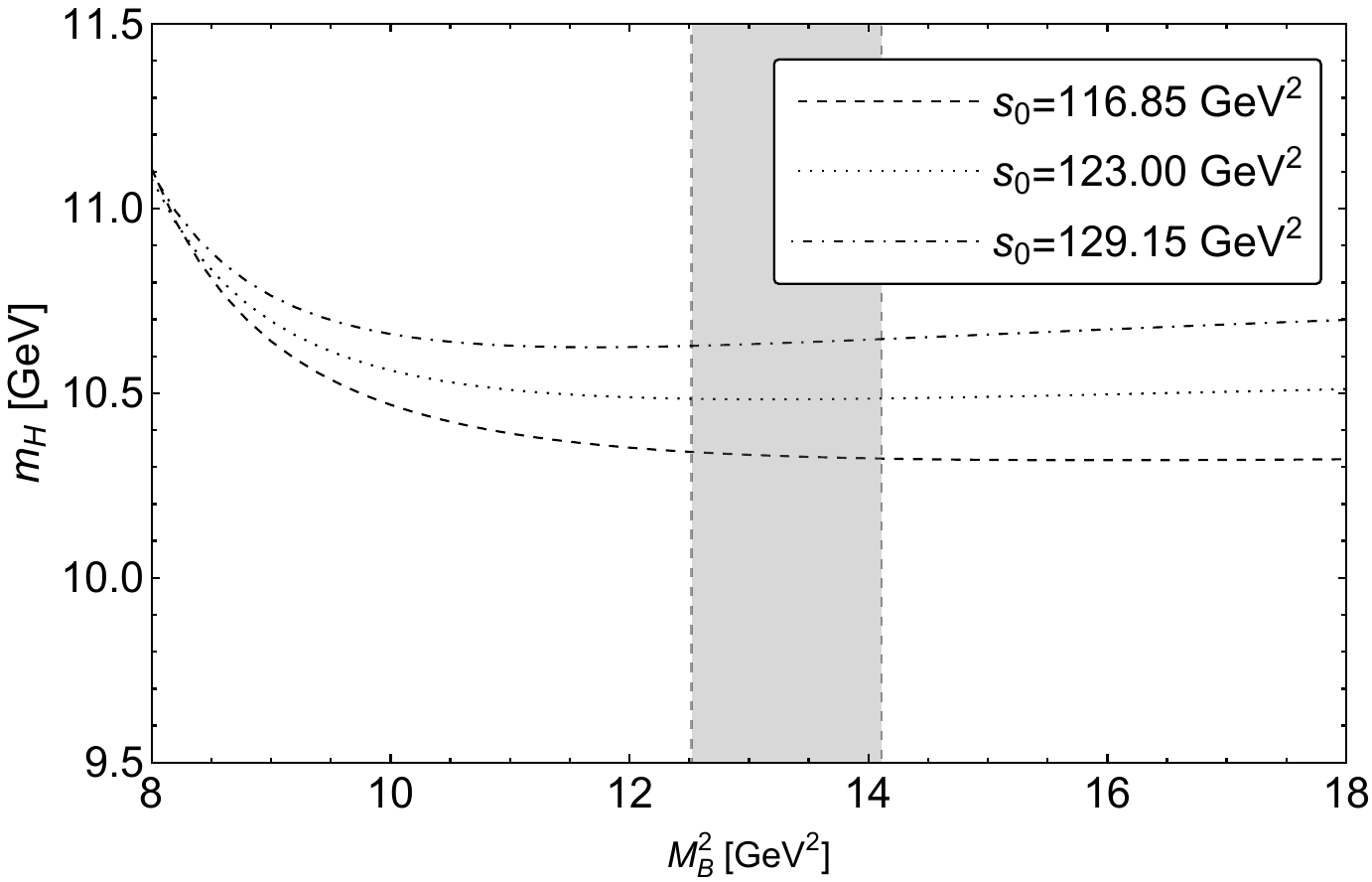}\label{fig:MB-mH-bGb}}\\
  \caption{Variation of $m_{H}$ with (a)$s_{0}$ and (b)$M_{B}^{2}$ corresponding to the $J^{PC}=2^{+-}$ $\bar{b}Gb$ hybrid state.}
\label{fig:Result-bGb}
\end{figure}

For $\bar{b}Gb$ hybrid state, similar numerical analysis can be performed. We show the OPE convergence and the Borel curves in Figs.~\ref{fig:Convergence-bGb} and \ref{fig:Result-bGb}, respectively. The obtained mass of $\bar{b}Gb$ hybrid state with $J^{PC}=2^{+-}$ is
\begin{align}
    m_{\bar{b}Gb}&=10.48_{-0.16}^{+0.16}~\mathrm{GeV}, \,
\end{align}
which is lower than the result from Ref.~\cite{Ryan:2020iog}

In Refs.~\cite{HadronSpectrum:2012gic,Dudek:2011bn}, the authors have studied the exotic hybrid mesons and found small mass splittings between $0^{+-}$ and $2^{+-}$ states. To investigate this relationship, we also study $J^{PC}=0^{+-}$ heavy quarkonium hybrid mesons. We obtain the projection operators and spectral functions and list them in the Appendix.~\ref{appendix2}. However, we find that the spectral functions for $J^{PC}=0^{+-}$ hybrid states obtained in this work are numerically same as those in Ref.~\cite{Chen:2013zia}, despite an overall multiplicative factor. The reason for this conclusion is that, in Ref.~\cite{Chen:2013zia}, the current used to extract the $0^{+-}$ hybrid state is $\bar{Q} g_s \gamma_\nu \gamma_5 \tilde{G}_{\mu \nu } Q$, which is just one contracted form of current $J_{\alpha\mu\nu}^{2}$ and can be projected out by introducing an appropriate projection operator with an overall factor remained.

\begin{figure}[h!!]
  \centering
  \subfigure[]{\includegraphics[width=4.2cm]{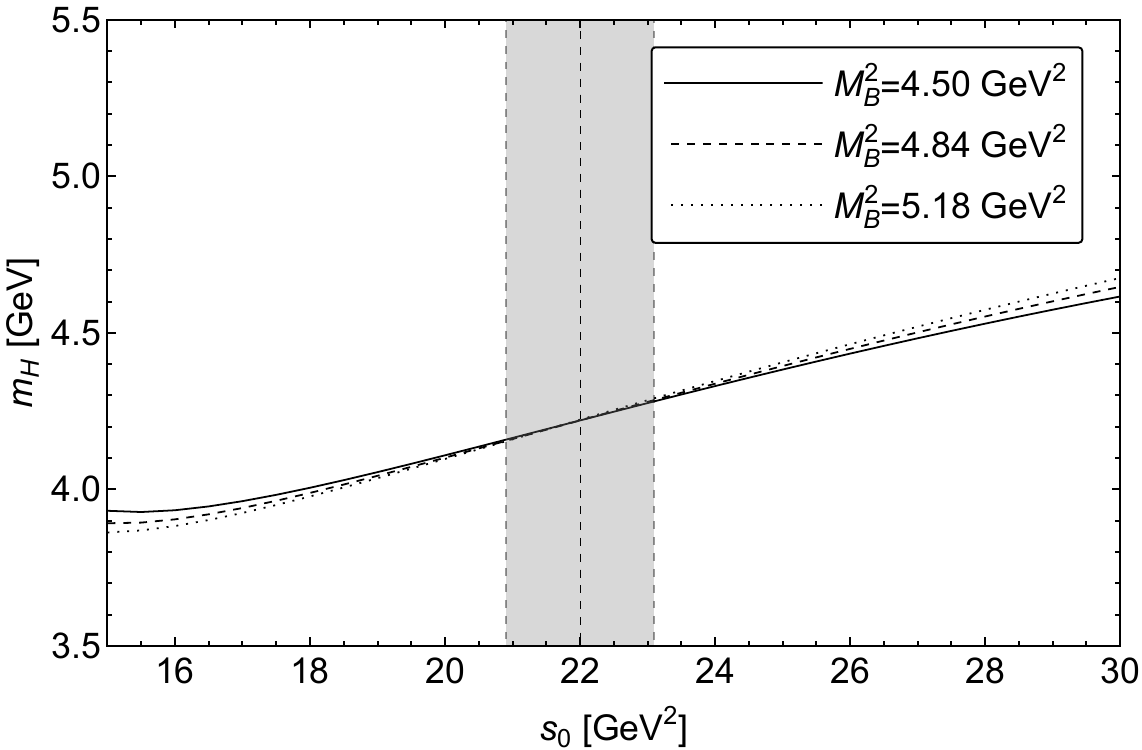}\label{fig:s0-mH-cGc0}}
  \subfigure[]{\includegraphics[width=4.2cm]{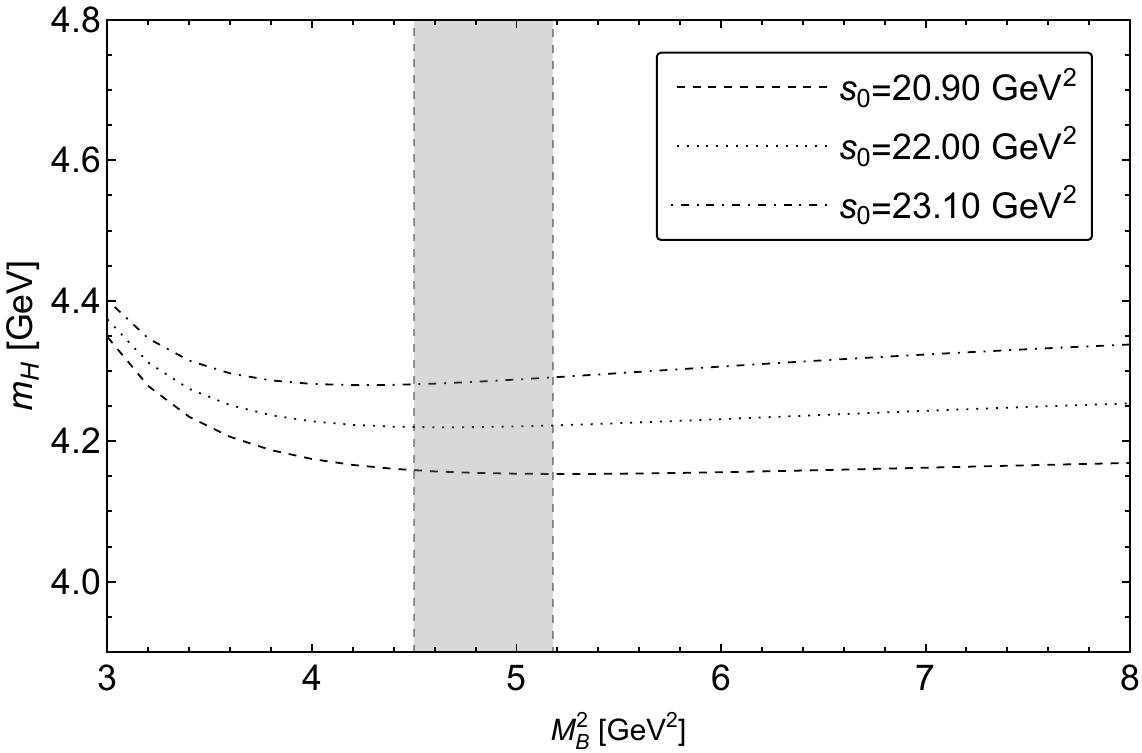}\label{fig:MB-mH-cGc0}}\\
  \caption{Variation of $m_{H}$ with (a)$s_{0}$ and (b)$M_{B}^{2}$ corresponding to the $J^{PC}=0^{+-}$ $\bar{c}Gc$ hybrid state.}
\label{fig:Result-cGc0}
\end{figure}

\begin{figure}[h!!]
  \centering
  \subfigure[]{\includegraphics[width=4.2cm]{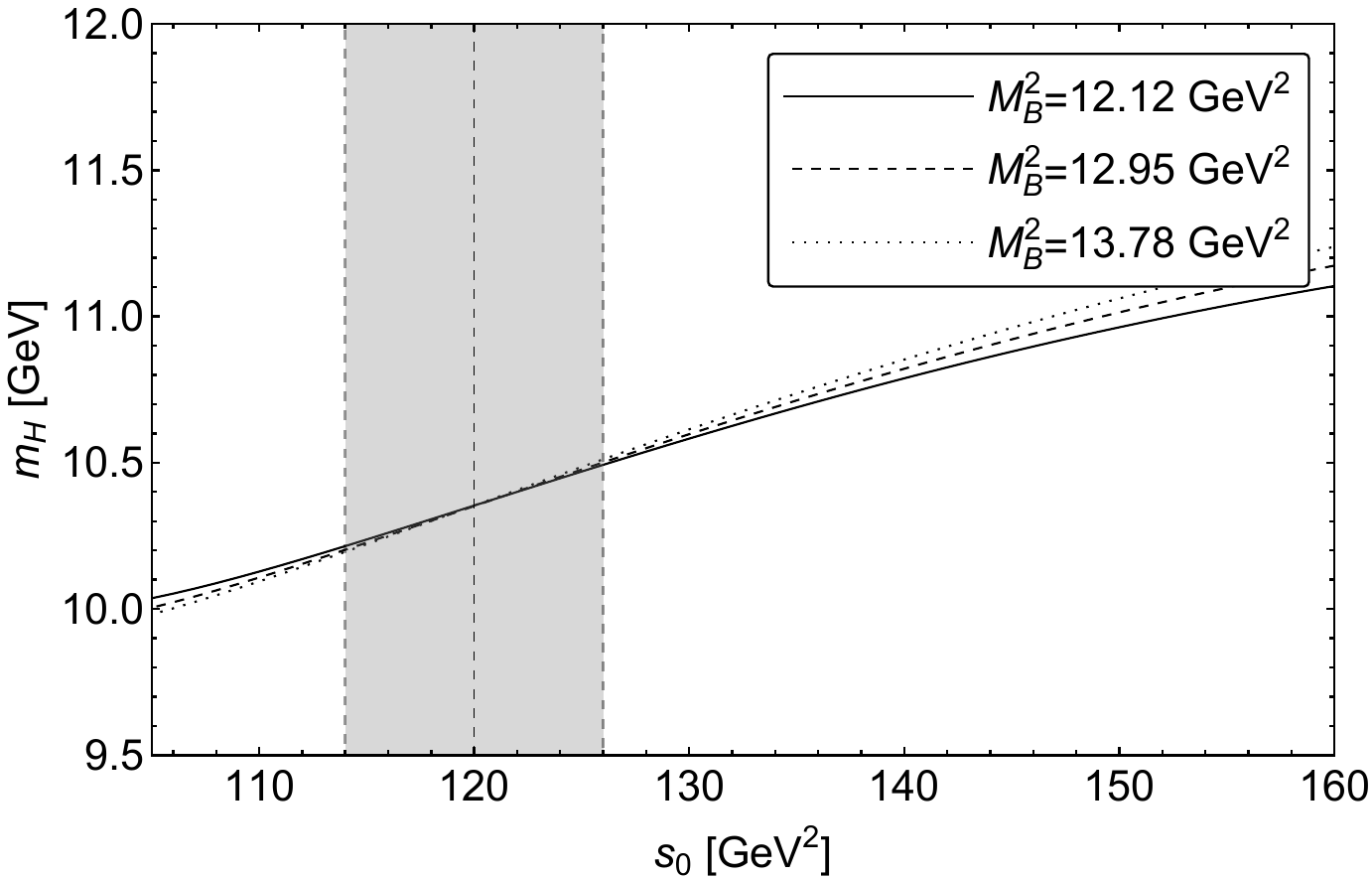}\label{fig:s0-mH-bGb0}}
  \subfigure[]{\includegraphics[width=4.2cm]{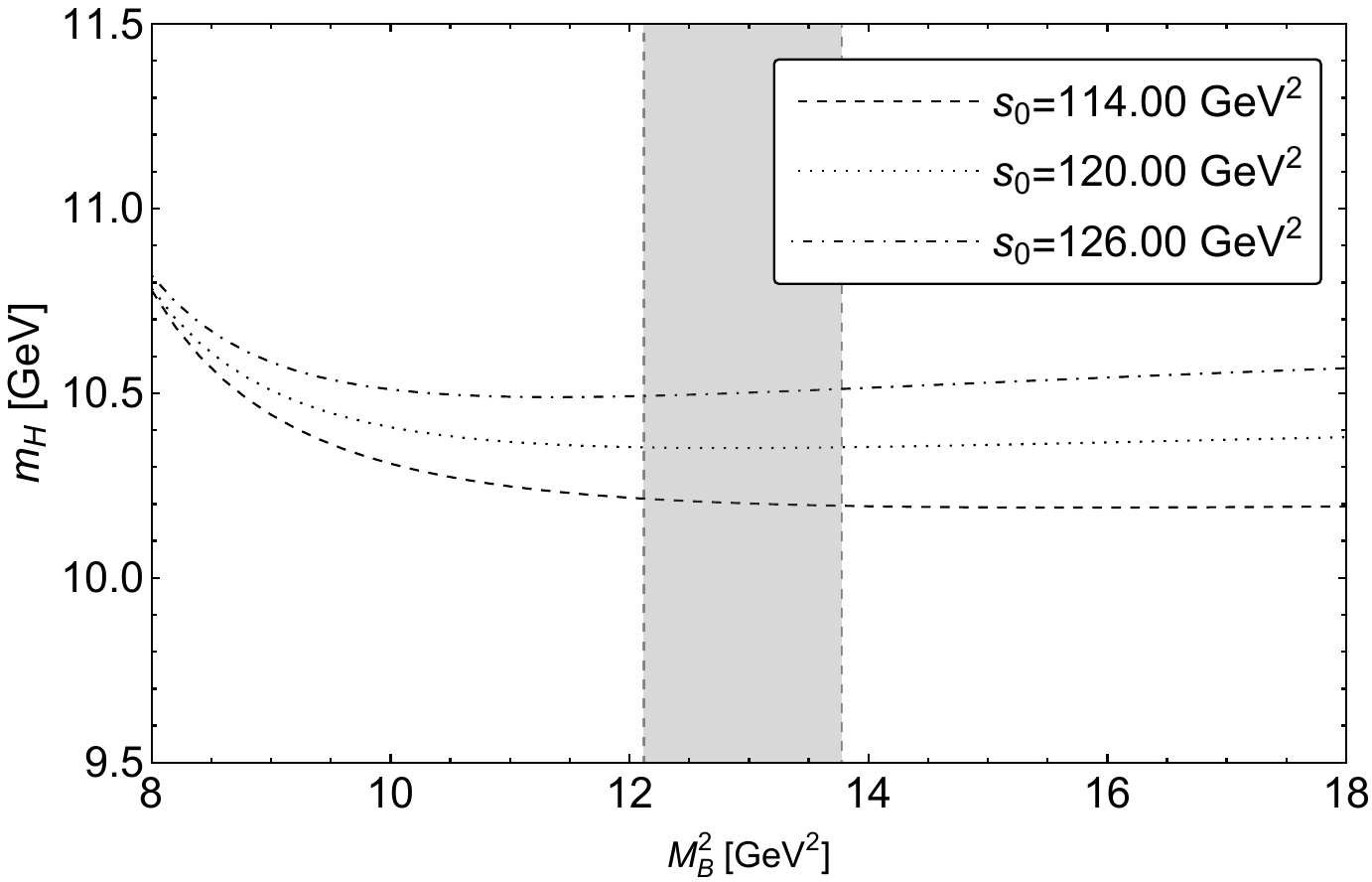}\label{fig:MB-mH-bGb0}}\\
  \caption{Variation of $m_{H}$ with (a)$s_{0}$ and (b)$M_{B}^{2}$ corresponding to the $J^{PC}=0^{+-}$ $\bar{b}Gb$ hybrid state.}
\label{fig:Result-bGb0}
\end{figure}

Besides, the values of condensates parameters applied in Ref.~\cite{Chen:2013zia} are different from this work. To study the mass splittings between $0^{+-}$ and $2^{+-}$ states, we reanalyse the masses of states with $J^{PC}=0^{+-}$ by performing the similar numerical analysis as those for the $2^{+-}$ states in this work. The Borel curves for $\bar{c}Gc$ and $\bar{b}Gb$ states with $J^{PC}=2^{+-}$ are shown in Figs.~\ref{fig:Result-cGc0} and \ref{fig:Result-bGb0}. We list all results in Table.~\ref{tab:results}, from which one can see that for $\bar{c}Gc$ and $\bar{b}Gb$, the masses of the $0^{+-}$ states are slightly lower than the masses of $2^{+-}$ states by approximately $100-300 ~\mathrm{MeV}$, which is roughly consistent with the conclusions in Ref.~\cite{HadronSpectrum:2012gic,Dudek:2011bn}. Since we adopt different condensate values, the masses obtained in the present work are about 5\% heavier than those reported in Ref.~\cite{Chen:2013zia}.


\begin{table}[t!]
  \caption{Mass spectra for heavy quarkonium hybrid mesons with $J^{PC}=(0,2)^{+-}$.}
\renewcommand\arraystretch{1.9} 
  \setlength{\tabcolsep}{0.5em}{ 
    \begin{tabular}{cccccc}
      \hline\hline  
      $\text{States}$ & $J^{PC}$  & $M_B^2[\mathrm{GeV}^2]$ & $s_0[\mathrm{GeV}^2]$ & $m_H$ [GeV] & $\text{P.C}[\%]$ \\
      \hline 
        \multirow{2}{*}{$\bar{c}Gc$} & $0^{+-}$  & 4.50-5.18 &  $22.0(\pm 5\%)$& $4.22_{-0.07}^{+0.07}$  & 34.7    \\
      & $2^{+-}$ & 5.21-5.61 &  $24.5(\pm 5\%)$ & $4.49_{-0.08}^{+0.08}$  & 32.4 \\   
      \cline { 2 -6 }\multirow{2}{*}{$\bar{b}Gb$ }& $0^{+-}$ & 12.12-13.78 &$120.0(\pm 5\%)$ &$10.35_{-0.15}^{+0.15}$  &34.6 \\
      &$2^{+-}$ & 12.52-14.11 &$123.0(\pm 5\%)$ &$10.48_{-0.16}^{+0.16}$  &34.2  \\ 
     \cline { 2 -6 }
      \hline\hline  
  \end{tabular}
}
\label{tab:results}
\end{table}


\section{Conclusion and Discussion}\label{Sec:5}
In our study, we utilized QCD sum rules to examine the mass spectra of heavy quarkonium  hybrid states $\bar{c}Gc$ and $\bar{b}Gb$ with exotic quantum numbers $J^{PC}=2^{+-}$. This analysis involved constructing interpolating currents with three Lorentz indices. We computed the correlation functions and spectral densities, taking into account condensates up to dimension 6. The states are extracted by constructing the corresponding projection operators. The obtained results give the masses of the $\bar{c}Gc$ and $\bar{b}Gb$ heavy quarkonium hybrid states, the corresponding masses are about $4.49~\mathrm{GeV}$ and $10.48~\mathrm{GeV}$, respectively, with a small mass splitting predicted between the $0^{+-}$ and $2^{+-}$ states.

The decay behavior of heavy quarkonium hybrid states with $J^{PC} = 2^{+-}$ is quite intriguing. One dominant mechanism proceeds as follows: a $\bar{q}q$ or $\bar{s}s$ quark pair is dynamically generated from the valence gluon field and subsequently fuses with the original valence quark--antiquark pair to form two separate mesons. For a $\bar{c}Gc$ hybrid with $J^{PC}=2^{+-}$, the most accessible strong decays are predicted to be into open-charm meson pairs, such as P-wave decays  $D\bar{D}_{1}$ and $D_{0}\bar{D}^{\ast}$, and D-wave decays like $D^{\ast}\bar{D}^{(\ast)}$ and $D_{s}^{\ast}\bar{D}_{s}^{(\ast)}$.

Furthermore, in heavy quarkonium systems, decay patterns are strongly constrained by heavy quark spin symmetry~\cite{Brambilla:2022hhi}. For $2^{+-}$ hybrids, the total spin of the heavy quarkonium component is $S(Q\bar{Q}) = 1$. This implies that decays into final states with $S(Q\bar{Q}) = 0$, which are spin-flipping processes, are highly suppressed, such as the P-wave decay $h_c \eta$, which remains suppressed even when considering possible enhancements from mechanisms like the QCD axial anomaly~\cite{Chen:2022qpd}. In contrast, spin-conserving decays remain allowed. These include P-wave decays such as $J/\psi f_{0,1,2}$ and $\chi_{c0,1,2} \omega$.

Tetraquark $[Qq][\bar{Q}\bar{q}]$ or hadronic molecules $[\bar{Q}q][\bar{q}Q]$ may also carry the quantum numbers $J^{PC} = 2^{+-}$. However, in such configurations, the heavy quark and antiquark reside in different diquark/antidiquark or mesonic clusters. Therefore, the spin-flipping decay into $h_c \eta$ is expected to be allowed, in contrast to the highly suppressed case for hybrid states. The branching ratios for spin-flipping versus spin-conserving decay channels of a hybrid state are expected to differ significantly from their counterparts in a tetraquark/molecular state. This striking difference makes decays such as $h_c \eta$ a key signature for distinguishing between hybrid and tetraquark/molecular interpretations.

We propose to search for $\bar{c}Gc$ hybrid meson with $J^{PC}=2^{+-}$ in $J/\psi f_{0,1,2}$ and $\chi_{c0,1,2} \omega$ final states. The $\bar{b}Gb$ hybrid state with $J^{PC}=2^{+-}$ could be relatively narrow due to limited phase space, which may restrict the number of accessible strong decay channels. Further investigations in both theories and experiments are needed to study these exotic states.

\section*{ACKNOWLEDGMENTS}
\sloppy
Qi-Nan Wang was supported by the the Natural Science Foundation of Liaoning Province of China under Grant No. 2025-BS-0816 and the Doctoral Startup Project of Bohai University under Grant No. 0525bs003. This work was also supported by the National Natural Science Foundation of China under Grant No. 12575153, 12175318 and 12075019.

\appendix
\section{Spectral functions of $2^{+-}$ states}\label{appendix1}
Spectral functions of $2^{+-} $ heavy quarkonium hybrid meson states are
\begin{align}
     \rho^{\text{pert}}&=\frac{\alpha _{s} }{48 \pi ^3} \int_{\alpha_{\min}}^{\alpha_{\max}} d\alpha \int_{\beta _{\min}}^{\beta _{\max}} d\beta 
      \frac{ (1-\alpha -\beta) F(\alpha ,\beta )^2 }{ \alpha ^2 \beta ^2} \nonumber\\
      &\times\left((1-3 \alpha -3 \beta ) F(\alpha ,\beta )+6 (\alpha +\beta ) \left(\alpha  \beta  s-m_{Q}^{2}\right)\right)\nonumber\\ 
      &=\frac{\alpha _{s} m_{Q}^6}{23040 \pi ^3 z^3} \left[15\log \left(\sqrt{z-1}+\sqrt{z}\right) \right.\nonumber\\
      & \times\left(1408 z^4-768 z^3+336 z^2-80 z+7\right)+\sqrt{z(z-1)}\nonumber\\
      &\left.\times\left(1280 z^5-9344 z^4-8752 z^3+4296 z^2-1130 z+105\right) \right], \,\\
  \rho^{GG}&= \frac{\dGG  \left(s-4 m_{Q}^{2}\right)}{72 \pi}\sqrt{1-\frac{4m_{Q}^{2}}{s}}\nonumber\\
    &=\frac{\dGG (z-1)^{3/2} m_{Q}^{2}}{18 \pi  \sqrt{z}}, \,\\
  \rho^{jj}&= \frac{\dJJ  \left(2m_{Q}^{2}+ s\right)}{192 \pi ^2 s}\sqrt{1-\frac{4 m_{Q}^{2}}{s}}+ \frac{ \dJJ m_{Q}^{4}}{36 \pi ^2 \sqrt{s^3 \left(s-4 m_{Q}^{2}\right)}}\nonumber\\
  &=\frac{\dJJ \left(6 z^2-3 z-1\right) }{1152 \pi ^2 \sqrt{z-1} z^{3/2}}, \,\\
  \rho^{G^3}&= \frac{\dGGG \left(4 m_{Q}^{2}+3 s\right)}{384 \pi ^2 s}\sqrt{1-\frac{4 m_{Q}^{2}}{s}} + \frac{\dGGG m_{Q}^{4}}{24\pi ^2 \sqrt{s^3 \left(s-4 m_{Q}^{2}\right)}}\nonumber\\
    &= \frac{\dGGG (3 z-2) }{384 \pi ^2 \sqrt{z(z-1)}}\,,
\end{align}
where
  \begin{align}
   & z=\frac{s}{4 m_{Q}^{2}} , \,\nonumber\\
   & \alpha_{\min }=\frac{1}{2}\left(1-\sqrt{1-\frac{4m_{Q}^{2}}{s}}\right)  , \,\nonumber\\
   & \alpha_{\max }=\frac{1}{2}\left(1+\sqrt{1-\frac{4m_{Q}^{2}}{s}}\right)  , \,\nonumber\\
   & \beta _{\min}=\frac{m_{Q}^{2}\alpha }{s\alpha -m_{Q}^{2}} ,\, \beta _{\max}=1-\alpha , \,\nonumber\\
   &  F(\alpha ,\beta )= m_{Q}^{2}(\alpha +\beta)- s \alpha \beta .
  \end{align}
\section{Projection operators and spectral functions of $0^{+-}$ states}\label{appendix2}
One can rewrite the coupling in Eq.~\eqref{Eq:coupling2} and \eqref{Eq:coupling1} as
\begin{align}
  \left\langle 0\left|J_{\alpha \mu \nu }^{1}\right| 0^{+-}(p)\right\rangle
  =&f_{0} \varepsilon_{\alpha \mu \nu \tau} p^\tau , \,
  \label{Eq:coupling21}
\end{align}
  \begin{align}
  \left\langle 0\left|J_{\alpha \mu \nu }^{2}\right| 0^{+-}(p)\right\rangle =&
  f_{0}^{\prime}(p_\nu g_{\alpha \mu }  - p_\mu g_{\alpha \nu })\, .
  \label{Eq:coupling81}
\end{align}
Thus, normalized projection operator for the  $0^{+-}$ states from current $J_{\alpha \mu \nu }^{1}$ is
\begin{align}
\mathbb{P}_{\alpha_{1}\mu_{1}\nu_{1},\alpha_{2}\mu_{2}\nu_{2}}^{0}=\frac{1}{36p^2}\varepsilon_{\alpha_{1} \mu_{1}  \nu_{1}  \tau_{1} } p^{\tau_{1}}
\varepsilon_{\alpha_{2}  \mu_{2} \nu_{2} \tau_{2}}  p^{\tau_{2}}  , \,
\label{Eq:Projector0a}
\end{align}
and from current $J_{\alpha \mu \nu }^{2}$ is
\begin{align}
  \mathbb{P}_{\alpha_{1}\mu_{1}\nu_{1},\alpha_{2}\mu_{2}\nu_{2}}^{0\prime}=&\frac{1}{36 p^2}( p_{\mu_{1}} g_{\alpha_{1} \nu_{1}}-p_{\nu_{1}} g_{\alpha_{1} \mu_{1}}) \nonumber\\
  &\times( p_{\mu_{2}} g_{\alpha_{2} \nu_{2}}-p_{\nu_{2}} g_{\alpha_{2} \mu_{2}})   \, .
  \label{Eq:Projector0b}
\end{align}

Spectral functions of $0^{+-} $ heavy quarkonium hybrid meson states are
\begin{align}
     \rho^{\text{pert}}&=\frac{\alpha _{s} }{72 \pi ^3} \int_{\alpha_{\min}}^{\alpha_{\max}} d\alpha \int_{\beta _{\min}}^{\beta _{\max}} d\beta 
      \frac{ (1-\alpha -\beta) F(\alpha ,\beta )^2 }{ \alpha ^2 \beta ^2} \nonumber\\
      &\times\left((2-3 \alpha -3 \beta ) F(\alpha ,\beta )+3 (\alpha +\beta ) \left(\alpha  \beta  s-m_{Q}^{2}\right)\right)\nonumber\\ 
      &=\frac{\alpha _{s} m_{Q}^6}{2160\pi ^3 z^3} \left[15\log \left(\sqrt{z-1}+\sqrt{z}\right) \right.\nonumber\\
      & \times\left(32 z^3- 16 z^2+ 6 z-1\right)+\sqrt{z(z-1)}\nonumber\\
      &\left.\times\left((32 z^4 - 224 z^3 - 188 z^2 + 80 z -15 )\right) \right], \,\\
  \rho^{GG}&= \frac{\dGG  \left(s-4 m_{Q}^{2}\right)}{216 \pi}\sqrt{1-\frac{4m_{Q}^{2}}{s}}\nonumber\\
  &=\frac{\dGG (z-1)^{3/2} m_{Q}^{2}}{54 \pi  \sqrt{z}}, \,\\
  \rho^{jj}&= \frac{\dJJ  \left(2 m_{Q}^{2}+ s\right)}{432\pi ^2 s}\sqrt{1-\frac{4 m_{Q}^{2}}{s}}+ \frac{\dJJ m_{Q}^{4}}{108\pi ^2 \sqrt{s^3 \left(s-4 m_{Q}^{2}\right)}}\nonumber\\
  &=\frac{\dJJ \left(4 z^2-2 z-1\right)}{1728 \pi ^2 z^{3/2}\sqrt{z-1} }, \,\\
    \rho^{G^3}&= \frac{\dGGG m_{Q}^{2}}{288 \pi ^2 s}\sqrt{1-\frac{4 m_{Q}^{2}}{s}} + \frac{\dGGG m_{Q}^{4}}{72 \pi ^2 \sqrt{s^3 \left(s-4 m_{Q}^{2}\right)}}\nonumber\\
    &= \frac{\dGGG }{1152\pi ^2 \sqrt{z(z-1)}}. \,
  \end{align}


\end{document}